\documentclass[12pt]{iopart}
\usepackage{graphicx}
\begin{document}
\title{Exact solution and asymptotic behaviour of the asymmetric simple exclusion process on a ring}
\author{Masahiro Kanai$^1$, Katsuhiro Nishinari$^2$ and Tetsuji Tokihiro}%
\address{$^1$Graduate School of Mathematical Sciences, 
University of Tokyo, 3-8-1 Komaba, Tokyo 153-8914, Japan}
\address{$^2$Department of Aeronautics and Astronautics, Faculty of Engineering, University of Tokyo, 7-3-1 Hongo, Tokyo 113-8656, Japan}
\ead{kanai@ms.u-tokyo.ac.jp}
\begin{abstract}
In this paper, we study an exact solution of the asymmetric
 simple exclusion process on a periodic lattice of finite sites
 with two typical updates, i.e., random and parallel.
Then, we find that the explicit formulas for the partition
 function and the average velocity are expressed by the Gauss
 hypergeometric function.
In order to obtain these results, we effectively exploit
 the recursion formula for the partition function for the
 zero-range process.
The zero-range process corresponds to the asymmetric simple
 exclusion process if one chooses the relevant hop rates of
 particles,
 and the recursion gives the partition function, in principle,
 for any finite system size.
Moreover, we reveal the asymptotic behaviour of the average
 velocity in the thermodynamic limit, expanding the formula
 as a series in system size.
\end{abstract}
\section{Introduction}
The asymmetric simple exclusion process (ASEP) is an exemplar
 of stochastic many-particle systems and provides fundamental
 models for various collective phenomena within the framework
 of nonequilibrium statistical mechanics
 \cite{SZ,Schutz00,Evans00,Schutz03}.
With some particular conditions, the ASEP shows a wide variety
 of nonequilibrium critical phenomena such as boundary-induced
 phase transitions \cite{Krug}, spontaneous symmetry breaking
 \cite{EFGM}, and phase separation \cite{EKKM,AHR,AHR2}.
As described here, the ASEP has been extensively studied both
 with periodic boundaries and with open boundaries, and at the
 same time one may consider several update rules such as
 random and parallel \cite{RSSS}.
In this paper, we focus on the ASEP on a lattice of
 finite sites with the periodic boundary condition and then
 consider a discrete-time evolution both with random dynamics
 and with parallel dynamics.
The ASEP with the periodic boundary condition is suitable for
 applying a theoretical approach for the first time.

In the ASEP, particles exclusively occupy sites on a lattice
 and hop from site to site in a definite direction with a
 constant probability, i.e., they move as if being driven
 by a biased field.
The particles have a short-range interaction due to the
 exclusive occupation (i.e. the hard-core exclusion).
However, as far as the parallel dynamics is concerned,
 it may induce a long-range interaction
 as the density of particles grows.
(Note that a parallel update rule treats all the sites
 equally, and the ASEP with parallel dynamics is hence
 regarded as a stochastic cellular automaton.)

Generally in stochastic models, an asymmetric dynamics causes
 asymmetric transition rates between states of configuration
 and it is manifested as a current of probability
 \cite{Blythe05}.
According to that current, one observes a macroscopic
 phenomenon such as a density flow, and the system is hence
 thought to be far from equilibrium without any equilibrium
 state.
However, nonequilibrium systems may have a steady state
 with a constant flow of particles, i.e., a nonequilibrium
 steady state \cite{EB}.
The ASEP and the zero-range process, both of which are the
 subjects of the present paper, are typical examples for that
 kind of nonequilibrium system.
We shall investigate the nonequilibrium steady state of these
 stochastic dynamical systems.

Analytic methods were improved especially in the context of
 transport phenomena, e.g., traffic flow, and then exact
 solutions were presented by powerful theoretical methods
 such as the matrix-product ansatz \cite{DEHP},
 and the cluster-approximation method \cite{SSNI}.
However, the previous results are for systems in the
 thermodynamic limit, i.e., of infinite system size.
As far as nonequilibrium many-particle systems are concerned,
 it is essential that one should investigate these systems
 of finite size
 and then consider them in the thermodynamic limit.
In this paper, we present an explicit formula of the
 partition function for the ASEP on a ring of finite sites,
 which allows for an exact calculation of physical values.

We find that the partition function is expressed by the Gauss
 hypergeometric function.
In order to obtain this result, we exploit the recursion
 formalism for the partition function for the zero-range
 process (ZRP) \cite{EH}, which is given in \cite{Evans97}.
As described later, the ASEP corresponds to the ZRP
 if one chooses the relevant hop rates of particles.
Since the partition function for the ZRP is, in principle,
 calculated for any finite system size via the recursion,
 one is able to obtain that for the ASEP in the same way.

As an application of the present result, we calculate an
 explicit formula for the average velocity of particles
 in the ASEP.
Due to the formulas for the hypergeometric function, we can
 provide an explicit formula for the average velocity by
 the Gauss hypergeometric function as well as
 for the partition function.
In addition to that, we provide the series expansion of the
 average velocity in system size by solving a sequence of
 equations of increasing order.
The series expansion reveals the asymptotic behaviour of
 a physical value in the thermodynamic limit.
In particular, we recover the well-known result given in
 \cite{SSNI}.

In most recent works \cite{Povolotsky,PM}, the same formulas
 as we obtain in the present work are given by using
 the Bethe ansatz solution of the master equation for the ZRP.
They find that the hop rates required by the applicability of
 the Bethe ansatz allow two parameters
 and in a special case the ZRP reduces to the ASEP.
It is interesting that, although the ZRP is concerned in both
 approaches, they seem entirely different until just
 before reaching the same formula.
We remark that, in contrast with the Bethe ansatz,
 our approach can be generalized to provide an exact solution
 of the ZRP if one takes other hop rates which do not comply
 with the conditions required by the Bethe ansatz \cite{inpre}.

This paper is organized as follows.
In section 2 we define the ZRP and see that in a special case
 the ZRP corresponds to the ASEP.
Then, we derive the partition function for the ASEP after preparing the general formalism for that for the ZRP.
In section 3 we provide the series expansion of the average
 velocity in system size.
Section 4 is devoted to conclusions and final remarks.
Long derivations of formulas are included in the appendix.
\section{Exact solution of the asymmetric simple exclusion process}
\subsection{The zero-range process}\label{seczrp}
The zero-range process is an exactly solvable stochastic
 process, and it has been widely used as a model for systems
 of many particles interacting through a short-range
 interaction \cite{EH}.
Particles in the ZRP are indistinguishable, occupying sites
 on a lattice (any lattice in any dimension), and each lattice
 site may contain an integer number of particles.
These particles hop to the next sites with a rate which depends
 on the number of particles at the departure site.
One is to apply the ZRP to phenomenological studies choosing
 a relevant hop rate of particles which determines the
 microscopic behaviour.

One of the most distinct properties of the ZRP is that its
 steady state is given by a factorized form \cite{EH}.
Each of these factors of the steady state are for each site of
 the lattice, and are determined by the hop rates.
In this sense, these factors are called the single-site weights.
We remark that the steady state behaviour is determined by
 the single-site weights and consequently one may infer the
 hop rates after choosing the single-site weights as desired.

The ZRP in one dimension can be mapped onto the ASEP if one
 rearranges particles and sites in the ZRP on a lattice in
 the ASEP in order that site $m$, containing $n_m$ particles
 in the ZRP, should be thought of as, in the ASEP, the $m$th
 particle occupying a site and unoccupied $n_m$ sites, i.e.,
 the distance to the $(m-1)$th particle in front of the
 $m$th particle.
In particular, the ZRP corresponds to the ASEP if one chooses
 the relevant hop rates (defined by (\ref{u})).
In recent works \cite{KNT,KNT2}, we proposed a traffic-flow
 model connecting the ASEP and the ZRP with an additional
 parameter.

In what follows, we consider the ZRP in one dimension
 with $N$ particles on a periodic lattice
 containing $M$ sites labelled $l=1,2,\ldots,M$.
It corresponds to the ASEP with $M$ particles
 on the periodic lattice of $L~(=M+N)$ sites,
 and then one should understand that the system is of size $L$
 and the density is $\rho=M/L$.

\subsection{The partition function for the ZRP}\label{secZ}
In this subsection, following \cite{EH}, we define the
 partition function for the ZRP and then find that the
 generating function for the partition function is obtained
 from that for the single-site weights.

The nonequilibrium steady state probability $P(\{n_m\})$ of
 finding the system in a configuration
 $\{n_m\}=\{n_1,n_2,\ldots,n_M\}$ is given as
 a product of the single-site weights denoted by $f(n)$:
\begin{equation}
P(\{n_m\})=\frac{1}{Z_{M,N}}\prod^M_{m=1}f(n_m)
\qquad(n_1+n_2+\cdots+n_M=N),
\end{equation}
 where a normalization $Z_{M,N}$
 is the so-called partition function.
The probability that a given site (e.g. site 1)
 contains $n$ particles is given by
\begin{equation}
p(n)=\sum_{n_2+n_3+\cdots+n_M=N-n}P(\{n,n_2,\ldots,n_M\})
= f(n)\frac{Z_{M-1,N-n}}{Z_{M,N}}.
\end{equation}
The sum of $p(n)$ over $n$ is unity by definition.
Thus, we obtain the recursion formula
 for the partition functions,
\begin{eqnarray}
Z_{M,N}&=&\sum^{N}_{n=0}f(n)Z_{M-1,N-n},\label{rec}\\
Z_{1,k}&=&f(k)\qquad(k\geq1).
\end{eqnarray}
Note that this is a recursion for a double series
 with respect to $M$ and $N$,
 and the partition functions are calculated
 recursively from the initial values $Z_{1,k}$.

Considering the generating functions
 $\widehat{f}(\zeta):=\sum^\infty_{n=0}f(n)\zeta^n$,
 and $\widehat{Z}_M(\zeta):=\sum^\infty_{n=0}Z_{M,n}\zeta^n$,
 we have a recursion with respect only to $M$:
 $\widehat{Z}_{M}(\zeta)=\widehat{f}(\zeta)\widehat{Z}_{M-1}(\zeta)$.
Accordingly, we find the fundamental relation
\begin{equation}
\widehat{Z}_M(\zeta)=\left(\widehat{f}(\zeta)\right)^M.
\label{genZ}
\end{equation}
Thus, the partition function $Z_{M,N}$ for any $N$ is
 obtained from the single-site weights $f(n)$.

The single-site weights, being obtained from the hop rates of
 particles, entirely characterize the zero-range process.
In \cite{EH,Evans97}, two formulas of the single-site weights
 respectively corresponding to random and parallel update rules
 are given as follows.
For a random update rule, the single-site weights $f(n)$ are
 expressed as
\begin{equation}
f(0)=1,\qquad f(n)=\prod^n_{j=1}\frac1{u(j)}\quad(n\geq1),
\label{rf}
\end{equation}
where $u(n)$ are the hop rates of particles when the departure
 site contains $n$ particles.
Note that $u(0)=0$ by definition.
For a parallel update rule, the single-site weights are
 expressed as
\begin{equation}
f(n)=\left\{
\begin{array}{ll}
1-u(1)&(n=0)\\
\displaystyle{
\frac{1-u(1)}{1-u(n)}\prod^n_{j=1}\frac{1-u(j)}{u(j)}}~&(n\geq1).
\end{array}
\right.
\label{pf}
\end{equation}

It is remarkable that according to the update rule
 the single-site weights have different recursions.
From (\ref{rf}) one finds that the single-site weights for
 a random update rule have a recursion
\begin{equation}
u(n)f(n)=f(n-1).\label{ruf}
\end{equation}
In a similar fashion, from (\ref{pf}) we have the recursion
 that the single-site weights for a parallel update rule
 satisfy,
\begin{equation}
u(n+1)f(n+1)=f(n)-u(n)f(n).\label{puf}
\end{equation}
We remark that as for the single-site weights the recursions
 (\ref{ruf}) and (\ref{puf}) may identify the different update
 rules instead of the explicit formulas (\ref{rf}) and
 (\ref{pf}), which will be seen in the course of calculations
 on the average velocity of particles in the ZRP.
\subsection{The average velocity of particles in the ZRP}
In this subsection, we shall provide the explicit formulas of
 the average velocity (or the mean hop rate averaged in the
 nonequilibrium steady state) of particles in the ZRP with
 random and parallel update rules.
In general, the flux of a transport system briefly presents its
 macroscopic property, and especially in nonequilibrium systems
 it is one of the few quantitative criteria.
In the present study, the flux of particles is obtained by
 multiplying the average velocity by the density of particles,
 since the number of particles is conserved due to the periodic
 boundary condition.

The average velocity, denoted by $v_{M,N}$, is defined by
\begin{equation}
v_{M,N}=\sum^{N}_{n=0}u(n)p(n)
=\sum^{N}_{n=0}u(n)f(n)\frac{Z_{M-1,N-n}}{Z_{M,N}}.\label{av}
\end{equation}
As mentioned in subsection \ref{secZ}, the recursion for the
 single-site weights plays the central role in the subsequent
 calculations.

We first consider the ZRP with the random rule.
From (\ref{rec}), (\ref{ruf}) and (\ref{av}), we find that the
 average velocity is expressed by the partition function, i.e.,
\begin{equation}
v_{M,N}=\frac{Z_{M,N-1}}{Z_{M,N}}.\label{rv}
\end{equation}

Next, we consider the parallel update rule.
From (\ref{rec}), (\ref{puf}) and (\ref{av}), we find the
 average velocity satisfies the recursion with respect to $N$,
\begin{equation}
v_{M,N+1}Z_{M,N+1}=Z_{M,N}-v_{M,N}Z_{M,N}.\label{vz}
\end{equation}
(It is suggestive that (\ref{vz}) is corresponding to
 (\ref{puf}).)
Solving (\ref{vz}) for $v_{M,N}$ with respect to $N$, the
 average velocity is expressed by the partition function, i.e.,
\begin{equation}
v_{M,N}=-\frac{\sum^{N-1}_{n=0}(-1)^nZ_{M,n}}{(-1)^NZ_{M,N}}.
\label{pv}
\end{equation}
\subsection{Exact solution of the ASEP}
Now, we turn to the calculations for the partition function and the average velocity in the ASEP.
As mentioned in subsection \ref{seczrp} one
 chooses the hop rates in the ZRP as
\begin{equation}
u(0)=0,\qquad u(n)=p\quad(0<p<1,\ n\geq1),\label{u}
\end{equation}
and the ZRP is thereby conformed to the ASEP with hop rate $p$.
\subsubsection{Random update rule}
First, we consider the ASEP with the random update rule.
From (\ref{rf}), the single-site weights are
 $f(n)=p^{-n}\ (n\geq0)$ and then the generating function
 for them becomes
\begin{equation}
\widehat{f}(\zeta)=\sum^\infty_{n=0}\left(\frac\zeta p\right)^n
=\frac1{1-\frac\zeta p}
\end{equation}
Hence, from (\ref{genZ}) the generating function
 for the partition functions is figured out as
\begin{equation}
\widehat{Z}_M(\zeta)=\left(1-\frac\zeta p\right)^{-M}
=\sum^\infty_{n=0}\frac{(M)_n}{p^nn!}\zeta^n.
\end{equation}
Thus, we obtain the partition function for the ASEP with the
 random update rule,
\begin{equation}
Z_{M,N}=\frac{(M)_N}{p^NN!}=
\frac1{p^{N}}\left(\!\!
\begin{array}{c}
M+N-1\\
M-1
\end{array}
\!\!\right),\label{rZ}
\end{equation}
where $(a)_n=a(a+1)\cdots(a+n-1)$ is the Pochhammer symbol.
Therefore, from (\ref{rv}) and (\ref{rZ}) the average
 velocity for the ASEP with the random update rule is given by,
\begin{equation}
v_{M,N}=\frac{Np}{M+N-1}.
\end{equation}
\subsubsection{Parallel update rule}
Next, we consider the ASEP with the parallel update rule.
From (\ref{genZ}) and (\ref{pf}), the generating function
 for the partition function becomes
\begin{equation}
\widehat{Z}_M(\zeta)
=\left(1-p+\frac{(1-p)\zeta}{p-(1-p)\zeta}\right)^M.\\
\label{pgenZ}
\end{equation}
Expanding (\ref{pgenZ}) in a power series of $\zeta$, we obtain
 the partition function for the ASEP with the parallel update rule,
\begin{eqnarray}
Z_{M,0}&=&(1-p)^M,\\
Z_{M,N}&=&\frac{(-1)^{N}(-p)^MM}{1-p}
F\!\left(\!\!
\begin{array}{c}
M+1,\,N+1\\
2
\end{array}
\!\!;\,\frac1{1-p}\right)
\qquad(N\geq1),\label{pZ}
\end{eqnarray}
 where
\begin{equation}
F\!\left(\!\!
\begin{array}{c}
\alpha,\,\beta\\
\gamma
\end{array}
\!\!;\,z\right)
=\sum^\infty_{n=0}\frac{(\alpha)_n(\beta)_n}{(\gamma)_n}\frac{z^n}{n!}
\end{equation}
 is the Gauss hypergeometric function.
(See \ref{ag} for details of the above calculations.)
This presentation allows one to take advantage of the formulas
 for the hypergeometric functions to make advanced calculations.

In order to calculate the average velocity according to
 (\ref{pv}), we first evaluate the sum in the numerator.
Using the Gauss recursion formula with respect to parameters
\begin{equation}
\frac{\alpha z}\gamma F\!\left(\!\!
\begin{array}{c}
\alpha+1,\,\beta+1\\
\gamma+1
\end{array}
\!\!;\,z\right)
=
F\!\left(\!\!
\begin{array}{c}
\alpha,\,\beta+1\\
\gamma
\end{array}
\!\!;\,z\right)
-
F\!\left(\!\!
\begin{array}{c}
\alpha,\,\beta\\
\gamma
\end{array}
\!\!;\,z\right),
\end{equation}
one finds
\begin{equation}
\sum^{N-1}_{n=0}(-1)^nZ_{M,n}=(-p)^M
F\!\left(\!\!
\begin{array}{c}
M,\,N\\
1
\end{array}
\!\!;\,\frac1{1-p}\right).
\end{equation}
Finally, the average velocity for the ASEP with the parallel update rule is obtained as
\begin{equation}
v_{M,N}=\frac{p-1}M
\frac{\displaystyle
F\!\left(\!\!
\begin{array}{c}
M,\,N\\
1
\end{array}
\!\!;\,\frac1{1-p}\right)}{
\displaystyle
F\!\left(\!\!
\begin{array}{c}
M+1,\,N+1\\
2
\end{array}
\!\!;\,\frac1{1-p}\right)}.
\label{1v}
\end{equation}
\section{Asymptotic behaviour of the exact solutions}
In this section, we investigate the asymptotic behaviour of the
 exact solutions of the ASEP, i.e., we find the power series
 expansion of the partition function $Z_{M,N}$ and the average
 velocity $v_{M,N}$, for a given density $\rho=M/L$, with
 respect to the system size $L~(=M+N)$.
As mentioned in the introduction, the average velocity is a
 physical value and is hence expanded in a power series with
 respect to the system size.
In contrast, the partition function shall be expanded as an
 asymptotic series.
\subsection{Random update rule}
First, we consider the random update rule.
After being expressed with the system size $L$, the average
 velocity (\ref{rv}) is easily expanded in a
 power series as follows.
\begin{equation}
v_{M,N}=\frac{p(1-\rho)L}{L-1}=p(1-\rho)(1+L^{-1}+L^{-2}+\cdots).
\label{rav}
\end{equation}
From (\ref{rav}), we have an expected consequence
\begin{equation}
\lim_{L\rightarrow\infty}v_{M,N}=p(1-\rho).
\end{equation}
\subsection{Parallel update}
Next, we consider the parallel update.
For the sake of convenience, we shall present the power
 expansion of the average velocity before turning to the
 partition function.
To begin with, we change the independent variable from $p$ to
 $z=1/(1-p)$.
Then, using the formula for the derivative of the hypergeometric
 function:
\begin{eqnarray}
\frac{\rmd^n}{\rmd z^n}\left[z^{\gamma-1}(1-z)^{\alpha+\beta-\gamma}
F\!\left(\!\!
\begin{array}{c}
\alpha,\,\beta\\
\gamma
\end{array}
\!\!;\,z\right)\right]\\
\qquad=
(\gamma-n)_nz^{\gamma-n-1}(1-z)^{\alpha+\beta-\gamma-n}
F\!\left(\!\!
\begin{array}{c}
\alpha-n,\,\beta-n\\
\gamma-n
\end{array}
\!\!;\,z\right),
\end{eqnarray}
 we can transform the rhs of (\ref{1v})
 into a logarithmic derivative:
\begin{equation}
v_{M,N}=\frac{z-1}{M}\frac{\rmd}{\rmd z}
\log\Bigl(
z(1-z)^{M+N}
F\!\left(\!\!
\begin{array}{c}
M+1,\,N+1\\
2
\end{array}
\!\!;\,z\right)
\Bigr).\label{log}
\end{equation}
The Gauss hypergeometric differential equation
 for the hypergeometric function in the argument
 of the logarithm in (\ref{log}) is given by
\begin{equation}
z(1-z)\frac{\rmd^2w}{\rmd z^2}+[2-(M+N+3)z]\frac{\rmd w}{\rmd z}
-(M+1)(N+1)w=0.
\end{equation}
Accordingly, the argument of the logarithm also satisfies a
 differential equation
\begin{equation}
\frac{\rmd^2w}{\rmd z^2}
+\frac{1-M-N}{z-1}\frac{\rmd w}{\rmd z}
+\frac{MN}{z(z-1)}w=0,
\end{equation}
and we thus find that the average velocity $v_{M,N}$ is the
 solution of a Riccati equation
\begin{equation}
\frac{p(p-1)}M\frac{\rmd}{\rmd p}v_{M,N}
=v^2_{M,N}-\left(1+\frac{N}{M}\right)v_{M,N}+\frac{Np}{M}.
\label{ricca}
\end{equation}
Expanding the average velocity
 as $v_{M,N}=v_0(\rho)+v_1(\rho)L^{-1}+v_2(\rho)L^{-2}+\cdots$,
 we separate the Riccati equation according to power of $L$:
\begin{eqnarray}
v^2_0-\frac1\rho v_0+\frac{p(1-\rho)}\rho=0,\\
p(p-1)\frac{\rmd}{\rmd p}v_{j-1}
=\sum_{k+l=j\atop k,l\geq0}\rho v_kv_l
-v_j\qquad(j\geq1).
\end{eqnarray}
This sequence of equations, solved in turn starting from $v_0$,
 give the power series expansion of the average velocity
 $v_{M,N}$ with respect to the system size $L$,
\begin{eqnarray}
v_{M,N}&=&\frac{1-\sqrt{1-4p\rho(1-\rho)}}{2\rho}
+\frac{(1-\rho)p(1-p)}{1-4p\rho(1-\rho)}L^{-1}\nonumber\\
&&\quad
+\frac{(1-\rho)p(1-p)
\Bigl[1-2p+p(3p+1)\rho(1-\rho)\Bigr]}
{\Bigl[1-4p\rho(1-\rho)\Bigr]^{5/2}}L^{-2}\nonumber\\
&&\qquad+\cdots.\label{ave}
\end{eqnarray}
In particular, we recover the well-known result for the average
 velocity in the thermodynamic limit \cite{SSNI}, i.e.,
\begin{equation}
\lim_{L\rightarrow\infty}v_{M,N}=\frac{1-\sqrt{1-4p\rho(1-\rho)}}{2\rho}.
\end{equation}
\section{Conclusion and remark}
In the present paper, we provide an exact solution of the ASEP
 on a periodic lattice with respect to two typical dynamics,
 i.e., random update and parallel update.
To begin with, we focus on the following two facts; First, the
 ZRP corresponds to the ASEP if one takes the hop rates of
 particles to be a constant in the ZRP.
Second, the partition function of the ZRP is exactly solvable
 in the sense that the partition function for any system size
 is obtained by a recursive calculation.
Then, by solving the recursion with the relevant hop rates,
 we are to obtain the partition function for the ASEP.

For the random dynamics, it is given by the product
 of a power of the hop rate and the binomial coefficient
 in which the number of sites and that of particles appear.
For the parallel dynamics, it is given
 by the Gauss hypergeometric function in which
 the number of sites and that of particles equally appear
 as the parameters.
It is remarkable that the hop rate becomes the independent
 variable.
Thus, the mathematical structure of the ASEP becomes clear
 from the viewpoint of special functions.

By using the above results, we calculate the average velocities
 and obtain their series expansions in system size.
The series expansions confirm the previous results given
 in the thermodynamic limit.
Note that, if necessary, one can obtain
 the higher-order correction terms for the average veloctiy
 in the parallel dynamics
 by solving the sequence of equations in order.

From the present results, it is expected that one can estimate
 the partition function for a ZRP with other hop rates than
 those we take in this paper.
This will be reported in the near future.
\ack
M Kanai would like to thank Kazuo Okamoto and Hidetaka Sakai
 for a helpful discussion on detailed calculations.

This work is partially supported by a grant for the 21st
 Century COE program ``A Base for New Developments of
 Mathematics into Science and Technology'' from the Ministry
 of Education, Culture, Sports, Science and Technology, Japan.
\appendix
\section{Calculation on the generating function for the ASEP}
\label{ag}
From the hop rates (\ref{u}), the single-site weights are
\begin{equation}
f(0)=1-p,\qquad f(n)=\left(\frac{1-p}p\right)^n\quad(n\geq1).
\end{equation}
For simplicity, we let $\gamma=1-p$ and $\beta=(1-p)/p$ hereafter.
Then, the generating function for the single-site weights is
\begin{equation}
\widehat{f}(\zeta)=\gamma+\sum^\infty_{n=1}(\beta\zeta)^n
=\gamma+\frac{\beta\zeta}{1-\beta\zeta}
\end{equation}
Accordingly, the generating function for the partition function
 becomes
\begin{equation}
\widehat{Z}_{M}(\zeta)=\left(\gamma+\frac{\beta\zeta}{1-\beta\zeta}\right)^M
=\sum^M_{k=0}\left(\!\!
\begin{array}{c}
M\\
k
\end{array}\!\!\right)
\gamma^{M-k}\left(\frac{\beta\zeta}{1-\beta\zeta}\right)^k.
\end{equation}
Using the Euler transformation for a series:
\begin{equation}
\sum^\infty_{k=0} a_kz^k
=\frac1{1+z}\sum^\infty_{n=0}\left(\frac{z}{1+z}\right)^nb_n
\quad
\mbox{where}
\quad
b_n:=\sum^n_{r=0}
\left(\!\!
\begin{array}{c}
n\\
r
\end{array}\!\!\right)a_r,
\end{equation}
we obtain
\begin{equation}
\widehat{Z}_{M}(\zeta)
=(1-\beta\zeta)\sum^\infty_{n=0}c_n(\beta\zeta)^n
=c_0+\sum^\infty_{n=1}(c_n-c_{n-1})(\beta\zeta)^n,\label{apgZ}
\end{equation}
where
\begin{equation}
c_n=\sum^n_{r=0}
\left(\!\!
\begin{array}{c}
n\\
r
\end{array}\!\!\right)
\!\left(\!\!
\begin{array}{c}
M\\
r
\end{array}\!\!\right)
\gamma^{M-r}.\label{cn}
\end{equation}
From (\ref{cn}), we have
\begin{eqnarray}
c_n-c_{n-1}&=&\sum^{n}_{r=1}
\left(\!\!
\begin{array}{c}
n-1\\
r-1
\end{array}\!\!\right)
\!\left(\!\!
\begin{array}{c}
M\\
r
\end{array}\!\!\right)
\gamma^{M-r}\\
&=&
M\sum^{N-1}_{r=0}\frac{(1-N)_r(1-M)_r}{(2)_r(1)_r}\gamma^{M-r-1}\\
&=&\gamma^{M-1}M
F\!\left(\!\!
\begin{array}{c}
1-M,\,1-N\\
2
\end{array}
\!\!;\,\gamma^{-1}\right).\label{cn2}
\end{eqnarray}
After all, we obtain (\ref{pgenZ}) from (\ref{apgZ}) and
 (\ref{cn2}).

Moreover, using the Kummer's transformation formula
\begin{equation}
F\!\left(\!\!
\begin{array}{c}
\alpha,\,\beta\\
\gamma
\end{array}
\!\!;\,z\right)
=(1-z)^{\gamma-\alpha-\beta}
F\!\left(\!\!
\begin{array}{c}
\gamma-\alpha,\,\gamma-\beta\\
\gamma
\end{array}
\!\!;\,z\right),
\end{equation}
one obtains (\ref{pgenZ}) from (\ref{pZ}).

\end{document}